\documentclass[prl,superscriptaddress,footinbib,twocolumn,floatfix]{revtex4}
\usepackage{graphics,epsfig,color,subfigure}
\usepackage{amssymb}
\usepackage{physics}
\makeatletter
\renewcommand{\@makecaption}[2]{% #1 is Figure Num, #2 is caption
  \vskip\abovecaptionskip
  \sbox\@tempboxa{\small\sf #1: #2}%
  \ifdim \wd\@tempboxa >\hsize
  \small\sf #1: #2\par
  \else
    \global \@minipagefalse
    \hb@xt@\hsize{\hfil\box\@tempboxa\hfil}%
  \fi
  \vskip\belowcaptionskip}
\makeatother

\def\be{\begin{eqnarray}}
\def\ee{\end{eqnarray}}

\def\Dslash{\,\,{\raise.15ex\hbox{/}\mkern-12mu D}}
\def\Dbarslash{\,\,{\raise.15ex\hbox{/}\mkern-12mu {\bar D}}}
\def\delslash{\,\,{\raise.15ex\hbox{/}\mkern-9mu \partial}}
\def\delbarslash{\,\,{\raise.15ex\hbox{/}\mkern-9mu {\bar\partial}}}
\def\pslash{\,\,{\raise.15ex\hbox{/}\mkern-9mu p}}
\def\calDslash{\,\,{\raise.15ex\hbox{/}\mkern-12mu {\cal D}}}

\def\lae{\mathrel{\mathop{\smash{\lower .5 ex \hbox{$\stackrel<\sim$}}}}}
\def\lae{\mathrel{\mathop{\smash{\lower .5 ex \hbox{$\stackrel>\sim$}}}}}

\newcommand{\Co}{{\bf C3}}

\newcommand{\CY}{{\bf CY}}

\newcommand{\Ho}{{\bf H}}
\newcommand{\Ket}[1]{{\bf [ #1\rangle} }
\newcommand{\SWAP}{{\bf SWAP}}
\newcommand{\Xo}{{\bf X}}
\newcommand{\Yo}{{\bf Y}}
\newcommand{\Zo}{{\bf Z}}

\begin{document}

\title{Quantum circuits with classically simulable operator scrambling}

\author{Mike Blake}
\author{Noah Linden}

\affiliation{School of Mathematics, University of Bristol, Fry Building, Woodland Road, Bristol BS8 1UG, UK}

\begin{abstract}
We introduce a new family of quantum circuits for which the scrambling of a subspace of non-local operators is classically simulable. We call these circuits `super-Clifford circuits', since the Heisenberg time evolution of these operators corresponds to  Clifford evolution in operator space. Thus we are able to classically simulate the time evolution of certain single Pauli strings into operators with  operator entanglement that grows linearly with the number of qubits. These circuits provide a new technique for studying scrambling in systems with a large number of qubits, and are an explicit counter example to the intuition that classical simulability implies the absence of scrambling.  
\end{abstract}
  %\cite{Nahum:2016muy}
%\keywords{}

\maketitle

%\pdfoutput=1
%\pagestyle{plain} \setcounter{page}{1}
%\newcounter{bean}
%\baselineskip16pt

\paragraph{Introduction.} 
Over the last few years many key, and intrinsically quantum, aspects of the dynamics of many-body quantum systems have been revealed by studying the Heisenberg time evolution of operators \cite{Nahum:2017yvy,Khemani:2017nda,Shenker1,Roberts,Roberts:2018mnp,Prosen,Swingle,Hosur:2015ylk,MSS,Liu:2019svk}. For  generic quantum dynamics one expects this time evolution to be `chaotic' - in the sense that an initially simple operator will eventually become scrambled amongst a large number of degrees of freedom of the quantum system. Tracking this chaotic time evolution of operators can be thought of as a fundamentally quantum problem, since following this evolution with respect to some basis of operators requires monitoring an exponential number of amplitudes \cite{Prosen,Roberts:2018mnp,Khemani:2017nda,Nahum:2017yvy}. Nevertheless, over the last few years there have emerged remarkable universalities and structure in this scrambling process. These include the discovery of a fundamental bound on chaos in systems with many local degrees of freedom \cite{MSS}, new insights into the AdS/CFT correspondence and black hole information\cite{MS,Jensen,MSY,Kitaev,Susskind:2018tei,Qi:2018bje,Lin:2019qwu,Susskind:2019ddc,Liu:2020gnp}, and surprising connections between scrambling and hydrodynamics in many-body quantum systems \cite{Gu,Patel,Aleiner,MB,MB4,Davison,Saso,Jonay:2018yei,vonKeyserlingk:2017dyr,Rakovszky:2017qit,Blake:2017ris,Blake:2018leo}.  

A closely related question occurs within quantum computation.   A generic quantum circuit creates time evolution leading to (entangled) states with exponentially many amplitudes being non-zero, starting from, say, the all-zero initial state.  The fact that there seems to be no way to keep track of these amplitudes on a classical computer, in general, with only polynomial effort, is fundamental to our belief that quantum computation is more powerful than classical computation.  
There are some families of circuits, however, notably Clifford circuits, which can be classically simulated \cite{aaronson, Gottesman:1998hu}.  Their dynamics can be computed classically with an overhead polynomial in the number of qubits by keeping track of the stabilizers of the states. Clifford dynamics, therefore, is considered classical, not being able to exhibit the full complexity of generic quantum evolution.

In this Letter we ask whether there are systems for which the chaotic time evolution of an operator can be computed efficiently using a classical computer. In particular, consider a many-body quantum system composed of $N$ qubits. Then there is a natural choice of initially simple operators, corresponding to operators that are single Pauli strings i.e. a product of Pauli matrices at each site. Under a generic time evolution such an operator will evolve into a linear superposition of Pauli strings and become an entangled state in the Hilbert space of operators (see e.g. \cite{Roberts:2018mnp,Prosen,Nahum:2017yvy,Khemani:2017nda}). Since the number of possible Pauli strings in this superposition grows exponentially with time,  one would not naively expect to be able to simulate this growth of operator entanglement classically. Consistent with this expectation is that under a Clifford evolution a single Pauli string remains a single Pauli string for all time, and hence never develops operator entanglement \cite{Nahum:2017yvy,vonKeyserlingk:2017dyr,Chen:2018hjf}. 

The purpose of this Letter is to introduce a  new family of quantum circuits, that we call `super-Clifford circuits', for which the scrambling of a subset of single Pauli strings can in fact be simulated classically. In particular we will present an explicit gate set for which the time evolution of the subspace of operators spanned by strings of $X$'s and $Y$'s can be regarded as a Clifford circuit in \emph{operator space}. Starting from a single such Pauli string, e.g. $O = X_1X_2 \dots X_N$, these `super-Clifford circuits' are capable of generating a near-maximal amount of operator entanglement within this subspace of operators in time polynomial in the number of qubits. Nevertheless we will show that the time evolution of $O$ can be simulated, also in polynomial time classically by tracking the evolution of a set of $N$ `super-stabilisers'. These super-stabilisers provide a highly efficient way of characterising the information contained in the operator wave-function, and can be used to explicitly compute the growth of operator entanglement in these circuits.

\paragraph{Gate set and dynamics in operator space.}
We begin by introducing our gate set and demonstrating that the time evolution of a subspace of operators under these gates can be regarded as a Clifford evolution in operator space.  We then use these gates to build an explicit deterministic circuit that generates an amount of operator entanglement linear in the number of qubits starting from a single Pauli string, before discussing generic circuits built from these gates.  

Our starting point is the phase gate $T$ in state space: $T\ket 0=\ket 0$, $T\ket 1=e^{i\pi/4}\ket 1$. It acts on operators (in the Heisenberg picture) as 
\begin{equation} 
T^\dagger X T = \frac{X - Y}{\sqrt 2},\quad T^\dagger YT = \frac{X + Y}{\sqrt 2}.
\label{T}
\end{equation}
The action \eqref{T} is rather like a Hadamard operator in the subspace of operators spanned by $X$ and $Y$.  To make this precise we adopt a state-like notation where $X$ is denoted $\Ket 0$ and $Y$ is denoted $\Ket 1$. Then the action of $T$ in state space induces an action on this subspace of operators that corresponds to the super-operator Hadamard $\Ho$ followed by the super-operator Pauli $\Zo$. I.e. the super-operator $\Zo. \Ho$ on this Hilbert space of operator strings acts as
\begin{eqnarray}
\Zo. \Ho \Ket 0 &=& \frac{\Ket 0 - \Ket 1}{\sqrt 2},\nonumber\\
\Zo. \Ho \Ket 1 &=& \frac{\Ket 0 + \Ket 1}{\sqrt 2},
\end{eqnarray}
corresponding to the action in \eqref{T}.  The super-operator $\Zo.\Ho$ is an example of a super-Clifford operator (i.e. an operator acting as a Clifford operator on this Hilbert space of operators). We denote super-operators in bold font, to distinguish them from operators in state space.

We now consider a system of $N$ qubits, and the subspace of operators spanned by Pauli strings consisting only of products of $X's$ and $Y's$ at each site. This defines a Hilbert space of operators of dimension $2^{N}$.  Our previous observation - that the $T$ operator in state space induces a Clifford operation in operator space - motivates us to ask whether we can find other gates in state space that induce interesting super-Clifford operators on this subspace of operators.

We were not aware of any {\em a priori} argument that it should be possible to find such gates. Furthermore, initial attempts to induce the simplest Clifford operations are discouraging: for example one cannot induce a controlled not operator in this subspace of operators.  Such a putative operator would act as follows:
\begin{eqnarray}
X_1X_2 &\mapsto& X_1X_2,\nonumber\\
Y_1X_2 &\mapsto& Y_1Y_2.
\label{action2}
\end{eqnarray}
However it is not possible for any unitary on state space to induce this action since \eqref{action2} does not preserve commutation relations.

Furthermore it is clear that when a typical gate in state space acts on strings of operators involving only $X$'s and $Y$'s, it generically takes us outside this subspace of operators (by introducing Pauli $Z$ or the identity operator at certain sites). Even amongst those gates whose action on operators does preserve the space of strings involving only $X$'s and $Y$'s, it is not clear that there are any such actions that correspond to other interesting super-Clifford operators (as opposed to more general operators on these strings).

Nevertheless, we now show that other interesting examples of induced super-Clifford operators do exist. Firstly we note that although we cannot induce controlled-not operators, we can induce the swap operator in our subspace of strings.  The 2-qubit swap gate  $SWAP$ in state space acts as $SWAP\ket \psi\ket \phi\ket= \ket \phi\ket \psi$ for arbitrary single qubit states $\ket \psi,\,\ket \phi$.  It induces a superoperator ${\SWAP}$: e.g.
\begin{equation}
\SWAP \Ket {01}= \Ket {10},
\end{equation}
corresponding to 
\begin{equation}
SWAP^\dagger X_1Y_2 SWAP = Y_1X_2.
\end{equation}

Furthermore, after some trial and error, we found the following gate, that we denote $C3$, which we will show acts in operator space as a product of controlled-$Y$ gates. This gate $C3$ is a three-qubit Clifford gate in state space constructed from the controlled-$X$ ($CX$)  and controlled-$Z$ ($CZ$) gates and $T$ as 
\begin{equation} 
C3=CX_{21} CX_{31} CZ_{12} T_1^{6} T_2^{6},
\end{equation}
where the notation $CX_{ab}$ means controlled-$X$ with $a$ as the control and $b$ as the target, and $T_a$ means the $T$ gate acting on qubit $a$. The gate $C3$ not only preserves strings of $X$'s and $Y$'s, but also acts as a Clifford operator when acting on such strings:
\begin{eqnarray}
C3^\dagger X_1X_2X_3 C3 &=& X_1X_2X_3,\nonumber\\
C3^\dagger X_1X_2Y_3 C3 &=& X_1X_2Y_3,\nonumber\\
C3^\dagger X_1Y_2X_3 C3 &=& X_1Y_2X_3,\nonumber\\
C3^\dagger X_1Y_2Y_3 C3 &=& X_1Y_2Y_3,\nonumber\\
C3^\dagger Y_1X_2X_3 C3 &=& -Y_1Y_2Y_3,\nonumber\\
C3^\dagger Y_1X_2Y_3 C3 &=& Y_1Y_2X_3,\nonumber\\
C3^\dagger Y_1Y_2X_3 C3 &=& Y_1X_2Y_3,\nonumber\\
C3^\dagger Y_1Y_2Y_3 C3 &=& -Y_1X_2X_3.
\label{C3}
\end{eqnarray} 
The action \eqref{C3} of the $C3$ operator in state space induces a super-operator $\Co$ that can be decomposed in terms of controlled-$\Yo$ super operators as $\Co=\CY_{12}\CY_{13} $. e.g.
\begin{eqnarray}
\Co \Ket {000} &=& \CY_{12}\CY_{13}\Ket {000} = \Ket {000},\nonumber\\
\Co \Ket {100} &=& \CY_{12}\CY_{13}\Ket {100} = -\Ket {111}, \quad {\rm etc.}
\end{eqnarray}

We are now able to make our key observation. We have constructed an explicit set of gates $ \{SWAP,T, C3\}$ that both preserve the operator space spanned by strings of $X's$ and $Y's$ and further act as Clifford operations $\{\SWAP,\Zo.\Ho, \Co\}$ within this operator space. It is therefore possible to classically simulate the time evolution of operators in these circuits by adapting the standard Clifford techniques to operator space.  Nevertheless these operations do not preserve the number of Pauli strings as the operator evolves, as can be seen explicitly from the action of $T$ in \eqref{T}.  
%\footnote{Note that our gates do not generate the entire Clifford group in operator space - as we have discussed,  the standard ${\bf CNOT}$ can never be induced by a unitary dynamics in state space.  Nevertheless the subgroup $\{\SWAP,\Zo.\Ho,\Co\}$ is sufficiently to generate operator entanglement starting from a single Pauli string}. 
In particular we will now demonstrate that this evolution can generate significant amounts of operator entanglement (i.e. an amount linear in $N$) starting from a single unentangled string e.g. $X_1X_2\ldots X_N \equiv \Ket {00\ldots 0}$.

\paragraph{A deterministic circuit.} In this example, we consider a case where $N$ is a multiple of 3; $N=3k$. We start with the operator string $O = X_1X_2\ldots X_N \equiv \Ket {00\ldots 0}$.  We now consider the operator obtained by evolving this string by acting with $\tau$ successive gates drawn as 
\begin{equation}
\label{circuit}
    O_\tau = U_1^{\dagger} \dots U_\tau^{\dagger} O U_\tau \dots U_1, 
\end{equation}
corresponding to Heisenberg evolution of $O$ under the circuit $U = U_\tau \dots U_1$. In operator space this evolution produces a state $\Ket \Psi  = {\bf U_1}... {\bf U_\tau} \Ket {00\ldots 0} $ where ${\bf U_i}$ is the super-operator corresponding to the gate $U_i$. Note that in operator space we must first act with the final gate ${\bf U_\tau}$, as evident from the structure of \eqref{circuit}. 

 Let us now give a (short) circuit using our gate set $ \{SWAP,T, C3\}$ that can generate an amount of entanglement linear in the number of qubits, $N$. We first act on each of the first $k$ qubits with $T$; which induces $ \Zo. \Ho \Ket 0 = \frac{\Ket 0 - \Ket 1}{\sqrt 2}$ on each of these first $k$ operators. We now have a linear superposition of $2^{k}$ Pauli strings, but there is no entanglement since the corresponding state in operator space is a product state. To generate operator entanglement between sites we now perform a $C3$ gate from the $j$th qubit in the first block of $k$ qubits with the targets at the $j$th position in the second and third blocks; i.e. $C3_{j,k+j,2k+j}$. This can be performed locally by swapping the qubits so that the targets are next to the control, performing $C3$, then swapping back. In total this requires us to act with ${\cal O}(N^2)$ gates. The resulting operator string has $k$ ebits of entanglement between the first block and the remaining blocks.  In fact the resulting operator string $ O =  \frac{1}{\sqrt{2}^k} \prod_{j=1}^{k} (X_jX_{k+j}X_{2k+j}+Y_jY_{k+j}Y_{2k+j})$ corresponds to $k$ copies of an operator GHZ state
 \begin{equation}
 \frac{\Ket {000}+ \Ket {111}}{\sqrt 2}.
\end{equation}
So indeed the resulting operator string has $k=N/3$ ebits of entanglement between each of the 3 blocks of $k$ qubits and the rest of the system. 

\paragraph{Generic super-Clifford circuits.}
The above example demonstrates that our super-Clifford gates can be used to build circuits which generate operator entanglement starting from a single unentangled string. For that particularly simple circuit we could track the evolution of the operator directly. For a generic super-Clifford circuit this will not be possible, but nevertheless the Clifford property means that the time evolution of the operator string $O = X_1 X_2 \dots X_N$ can be simulated classically in operator space by tracking the time evolution of `super-stabilisers'. Furthermore tracking these super-stabilisers allows us to compute the evolution of operator entanglement in generic circuits built from our gate set. 

Let us therefore introduce the notion of super-stabilisers in the operator space spanned by strings of $X$'s and $Y$'s on $N$ qubits. To do this note that we can uniquely define an operator in this subspace by specifying $N$ independent super-operators ${\bf O}_{\alpha}$ under which the associated state $\Ket \Psi$ is invariant %
\begin{equation}
\label{stab}
{\bf O}_{\alpha} \Ket \Psi = \Ket \Psi .
\end{equation}
For our initial state corresponding to the operator $X_1X_2\ldots X_N \equiv \Ket {00\ldots 0}$ a set of these `super-stabilisers' is provided by the super-operators ${\bf Z}_{\alpha}$ for $\alpha = 1,\dots,N$. Under time evolution then equation \eqref{stab} still holds but with the stabilisers replaced by ${\bf O^\tau_{\alpha}}= {\bf U_1} ... {\bf U_\tau} {\bf O}_{\alpha} {\bf U_\tau^{\dagger}}... {\bf U_1^{\dagger}}$. Now since the operator-space time evolution is that of a super-Clifford circuit we know that super-stabilisers consisting of a single super-Pauli string (e.g. the ${\bf Z}_{\alpha}$ corresponding to our all $X$ string) remain a single super-Pauli string for all time. Hence they are of the form 
\begin{equation}
\label{decomposition}
{\bf O}^{\bf \tau}_{\alpha} \propto {\bf X}_{1}^{{\bf v}_{1x}}{\bf Z}_{1}^{{\bf v}_{1z}} \dots {\bf X}_{N}^{{\bf v}_{Nx}}{\bf Z}_{N}^{{\bf v}_{Nz}},
\end{equation}
up to a possible minus sign which we ignore since it does not effect the entanglement properties of the corresponding state in operator space. Each of our $N$ super stabilisers can thus be specified by a binary vector ${\bf v}$, where
\begin{equation}
{\bf v} = ({\bf v}_{1x}, {\bf v}_{1z},\dots,{\bf v}_{Nx},{\bf v}_{Nz}).
\label{vector}
\end{equation}

To track the time evolution of these super-stabilisers we just need to determine how our gates act on these stabiliser vectors ${\bf v}$. First let's start with the action of $T$. In operator space this induces the action of ${\bf Z}.{\bf H}$. To determine the action on the stabilisers \eqref{decomposition} it is sufficient to act on ${\bf X}$ and ${\bf Z}$ by this matrix. This gives 
\begin{equation}{\bf X} \to {\bf Z}, \;\;\; {\bf Z} \to -{\bf X}.
\end{equation}
Acting on the $i$th qubit therefore simply exchanges ${\bf v}_{ix}$ and ${\bf v}_{iz}$ in our vector \eqref{vector}.

We now want to determine the action of $\Co=\CY_{12}\CY_{13}$. When acting on the space of super-operators it acts by conjugation, e.g.  as $\Xo_1 \to \Co \Xo_1 \Co^\dagger=\Xo_1 \Yo_2\Yo_3 = - \Xo_1 \Xo_2 \Zo_2 \Xo_3 \Zo_3$. Acting on the rest of a basis for \eqref{decomposition} we find
\begin{eqnarray} 
 \Zo_1 \to  \Zo_1, \;\;   \Zo_2 \to \Zo_1 \Zo_2,  \;\; \Zo_3\to  \Zo_1\Zo_3,\nonumber \\ 
 \Xo_2 \to \Zo_1\Xo_2, \;\; \Xo_3 \to \Zo_1 \Xo_3. 
\end{eqnarray}
These transformations update ${\bf v}$ by 
\begin{eqnarray}
\label{controln}
{\bf v}_{1x} &\to & {\bf v}_{1x}, \;\; {\bf v}_{1z} \to {\bf v}_{1z} + {\bf v}_{2x}+{\bf v}_{2z} + {\bf v}_{3x}+{\bf v}_{3z} ,\nonumber\\
{\bf v}_{2x} &\to & {\bf v}_{1x} + {\bf v}_{2x} ,\;\; {\bf v}_{2z} \to {\bf v}_{1x}+ {\bf v}_{2z},\nonumber\\
{\bf v}_{3x} &\to & {\bf v}_{1x} + {\bf v}_{3x} ,\;\; {\bf v}_{3z} \to {\bf v}_{1x} + {\bf v}_{3z}.
\end{eqnarray}
Likewise $\SWAP$ acts by exchanging the components for the two qubits on which it acts. 

By updating the vector \eqref{vector} under such operations we are therefore able to simulate the evolution of operators by keeping track of $N$ stabilisers - i.e. a polynomial number of degrees of freedom. This happens even though under our dynamics the initial operator $X_1 X_2 \dots X_N$ has evolved into a sum of exponentially many operator strings - the point is that the super-stabilisers provide a highly efficient way of tracking this complicated time evolution. Further, following the techniques outlined in \cite{Nahum:2016muy} we can use them to compute the operator entanglement developed under this time evolution. To do this one considers the matrix formed by combining our $N$ super-stabilisers into a $2 N$ by $N$ matrix ${\bf V} = ({\bf v_1}^{T}, \dots, {\bf v_N}^{T})$. In terms of this matrix of stabilisers then the operator entanglement entropy of a subregion $A$ consisting of the first $p$ qubits is given by $S_A(t) = I_A - p$, where $I_A$ is the rank (in arithmetic modulo 2) of the submatrix formed by keeping the first $2 p$ rows of ${\bf V}$.

\paragraph{A random circuit.} As an explicit demonstration we now compute the time evolution of the operator entanglement in a particularly simple random circuit built from our gate set, starting from the single unentangled string  $X_1X_2\ldots X_N \equiv \Ket {00\ldots 0}$. For this state we chose an initial set of stabiliser vectors ${\bf v}_{\bf \alpha}$ with components ${\bf v}_{{\bf \alpha}, {ix}} = 0$ and ${\bf v}_{{\bf \alpha}, {iz}} = \delta_{ \alpha i}$. We now update these vectors under the action of a random circuit built by alternatively applying $T$ and $C3$ gates (in this random circuit we will not need to include $SWAP$ or higher powers of $T$ to generate near maximal operator entanglement). Our specific choice of random circuit is defined by first drawing a qubit $i=1,..,N$ and then acting with the gate $T$ on this qubit. Next we randomly draw a qubit with $i=1,...N-2$ and act with a $C3$ gate on qubits $i,i+1,i+2$. Since the action of $\Co$ is asymmetrical on the three qubits we also randomise which of these qubits is our control. We refer to this process of acting with a random $T$ gate and random $C3$ gate as one time step, and repeat over many time steps. At each time step we use the stabilisers to compute the operator entanglement of a subregion $A$ as explained above. 

\begin{figure}[!ht]
\includegraphics[width=60mm]{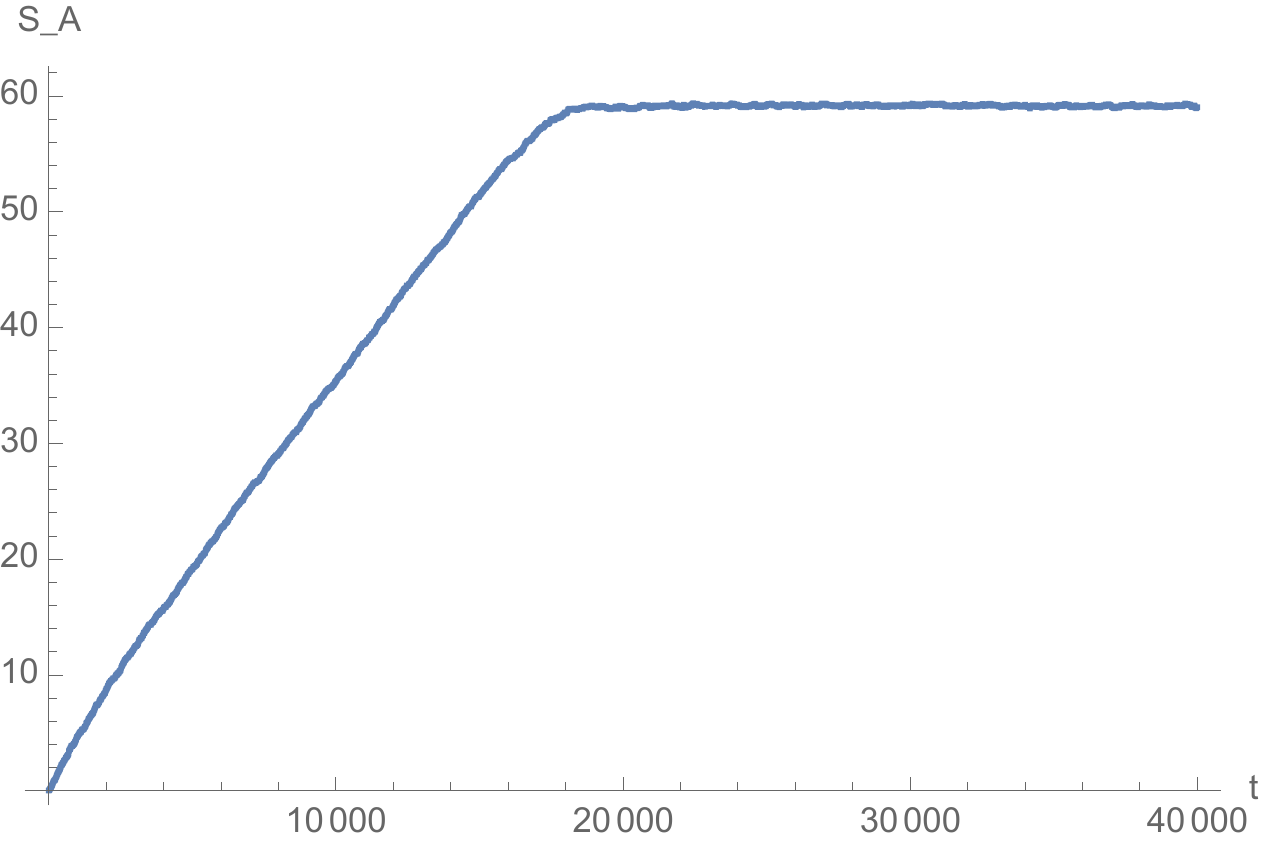}
\caption{The operator entanglement entropy (with base-2 logarithm) for an equal bipartition of a chain of $N=120$ qubits starting from the unentangled operator $X_1\dots X_N$. This plot is averaged over 50 realisations and $t$ counts the number of time steps.}
\label{entropy1}
\end{figure}

In Figure~\ref{entropy1} we illustrate the generic behaviour of operator entanglement in such circuits by computing the operator entanglement for an equal bipartition of a chain consisting of 120 qubits, averaged over 50 realisations to smooth out fluctuations. Initially the operator $X_1 X_2 \dots X_N$ is unentangled and hence the entropy is zero. However as we evolve under our circuit the operator entanglement increases linearly and then approximately saturates at a near maximal value that we find to be slightly less than the Page value \cite{Page:1993df}. As we vary the number of qubits we find that the rate of operator entanglement growth for the bipartition scales as ${\cal O}(1/ N)$ which can be understood heuristically from the fact that the probability of drawing a gate that acts at the cut scales as ${\cal O}(1/ N)$. In total the time taken for the operator entanglement to saturate therefore scales as ${\cal O}({N}^2)$.

\paragraph{Discussion.}

A natural question is whether we can extend our techniques to describe scrambling of local operators for this model. To simulate the circuits introduced in this Letter we considered only operators which had support on every site, i.e. those in the operator subspace spanned by strings containing only $X$'s and $Y$'s. Such operators are analogous to the homogeneous operators which were studied in the initial description of black hole scrambling  \cite{Shenker1}. However it is natural to ask about the scrambling and spreading of local operators, i.e. strings containing factors of the identity e.g. $I_1I_2\ldots X_j\dots I_N$. For the circuits discussed in this model it is simple to see that such operators can indeed scramble \footnote{For example acting with $C3$ with the $j$-th qubit the second target changes this string to one with $X$ at positions $j$ and $j+1$; thus continuing this  combined with $SWAP$s leads to a string with many $X$'s which will then scramble by the arguments earlier in this Letter.}, however we do not yet know of a way to classically simulate their operator dynamics for all circuits made from $\{SWAP, T, C3\}$. 

It would be interesting to understand the particular properties of the circuits we have considered here that enabled the simulation (we note, for example that our circuits preserve the computation basis \footnote{Operator dynamics in families of circuits with this property has been considered previously, for example, in the context of quantum automata \cite{sarang}.}).

More generally, extending the class of operators whose scrambling can be simulated classically is  a major question for  future research. The existence of such classical simulations of scrambling, such as those introduced in this paper, as well as providing insights into the nature of scrambling, should provide powerful new numerical methods for studying it. In particular it has the potential to allow us to explore scrambling and operator growth in systems with a large number of qubits, which has proved extremely challenging using existing techniques \cite{Kobrin:2020xms}. 

\section*{Acknowledgements} We are grateful to Adam Nahum for valuable conversations. NL gratefully acknowledges support from the UK Engineering and Physical Sciences Research Council through grants EP/R043957/1, EP/S005021/1, EP/T001062/1.

\end{document}